\documentclass[prl, aps, english, twocolumn, hyperref,floatfix,showpacs]{revtex4}
\usepackage{graphicx}
\usepackage{amssymb, amsmath, amsfonts, color, rotating, multirow, graphicx, bm}
\usepackage[bookmarks=false,pdfstartview=FitH,hyperindex=true, colorlinks, linkcolor=blue, citecolor=blue]{hyperref}

\usepackage{graphicx}

\begin{document}

\title{Theory of Bosons in two-leg ladders with large magnetic fields}

\author{Ran Wei and Erich J. Mueller}

\affiliation{Laboratory of Atomic and Solid State Physics, Cornell University, Ithaca, New York 14853}

\date{\today}

\begin{abstract}
We calculate the ground state of a Bose gas trapped on a two-leg ladder where Raman-induced hopping mimics the effect of a large magnetic field.  In the mean-field limit, where there are large numbers of particles per site, this maps onto a uniformly frustrated two-leg ladder classical spin model.  The net particle current always vanishes in the ground state, but generically there is a finite ``chiral current", corresponding to equal and opposite flow on the two legs.  We vary the strength of the hopping across the rungs of the ladder and the interaction between the bosons.  We find three phases: (1) A ``saturated chiral current phase" (SCCP), where the density is uniform and the chiral current is simply related to the strength of the magnetic field.  In this state the only broken symmetry is the $U(1)$ condensate phase.  (2) A ``biased ladder phase" (BLP), where the density is higher on one leg than the other.  The fluid velocity is higher on the lower density leg, so the net current is zero.  In addition to the $U(1)$ condensate phase, this has a broken $Z_2$ reflection symmetry.  (3) A ``modulated density phase" (MDP), where the atomic density is modulated along the ladder.  In addition to the $U(1)$ condensate phase, this has a second broken $U(1)$ symmetry corresponding to translations of the density wave. We further study the fluctuations of the condensate in the BLP, finding a roton-maxon like excitation spectrum. Decreasing the hopping along  the rungs softens the spectrum. As the energy of the ``roton" reaches to zero, the BLP becomes unstable. We describe the experimental signatures of these phases, including the response to changing the frequency of the Raman transition.
\end{abstract}
 \maketitle

\emph{Introduction ---}
The study of condensed bosons under rotation is an important and rich problem:
rotation probes superfluidity \cite{Hall1956} just like magnetic fields probe superconductivity \cite{Abrikosov1957}.
Such systems can be mapped onto a frustrated XY spin model \cite{Teitel1983},
and for large frustration and sufficiently large on-site interactions one finds the bosonic versions of 
the fractional quantum Hall effect \cite{Cooper2001,Cooper2009,Simon2012,Dalibard2013}. 
In the weakly interacting limit there are a rich variety of vortex phases \cite{Cooper2008}.
Here we study a Bose gas trapped on a two-leg ladder where Raman-induced hopping mimics the effect of a large magnetic field.

The bosonic two-leg ladder is appealing, as it is the simplest model for studying the response of bosons to a magnetic field.
Thus the experimental observations are particularly easy to interpret. Further, the ladder geometry is straightforward to model, admitting approaches
ranging from the density matrix renormalization group \cite{White1996} through bosonization \cite{Schulz1996}.
In the strongly interacting limit, there is an interesting interplay between Mott physics and the single particle band structure
\cite{Niccoli2008,Shin2011,Alex2013,Dhar2012,Dhar2013,Georges2014}. Here we use a mean-field analysis, which is appropriate
for describing experiments on arrays of weakly coupled ladders
when the number of particles per site is large.

Experimentalists in Munich have recently engineered this model \cite{Bloch2014}. Their
technique builds upon work performed at NIST, where Raman lasers created artificial magnetic fields in
the absence of a lattice \cite{Spielman2009}. Bloch's group generalized this idea and produced a staggered magnetic flux on an optical lattice \cite{Bloch2011}.
Later, both the Munich and MIT groups extended this to uniform fields \cite{Bloch2013,Ketterle2013}.
Other approaches to producing artificial gauge fields are reviewed by Dalibard \emph{et al}. \cite{Dalibard2011}.

\begin{figure}[!htb]
\includegraphics[width=8cm]{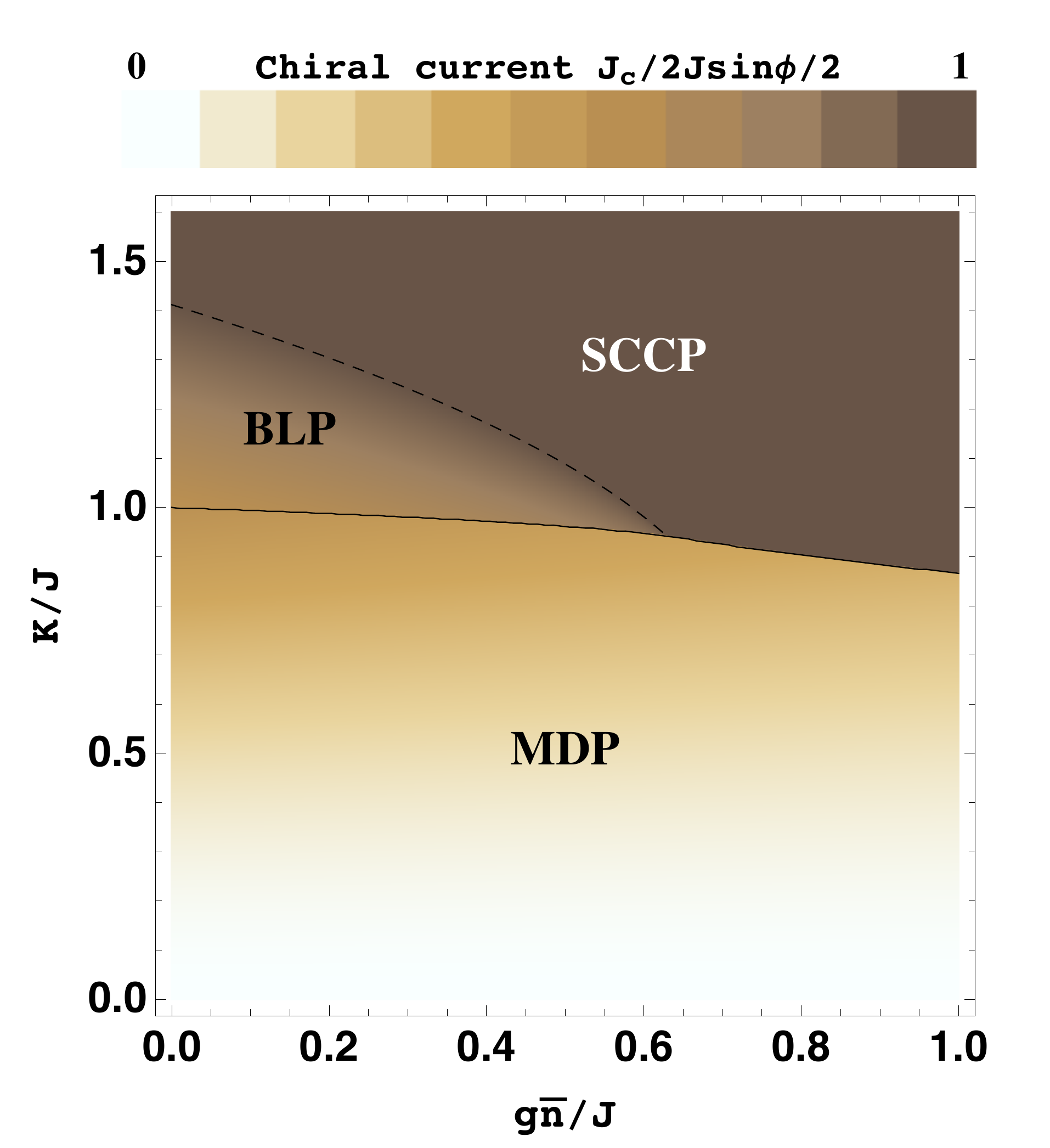}
\caption{(Color online) 
Phase diagram of a two-leg bosonic ladder as a function of the tunneling strength $K$ between the legs 
and interaction strength $g\bar n$ for a fixed flux per plaquette $\phi=\pi/2$. These energies are measured in terms of the strength of tunneling along the legs, $J$.
There are three phases: the ``saturated chiral current phase" (SSCP),  the ``biased ladder phase" (BLP) and the ``modulated density phase" (MDP).
The transition at the solid line is first-order, and the transition at the dashed line is second-order. 
The color represents the magnitude of the chiral current described by Eq. (\ref{chiraleq}).
Darker colors correspond to larger currents. The current is constant in the SSCP but varies in the BLP and MDP.
\label{phasediagram}}
\end{figure}

In this work, we use a variational approach to analytically calculate the ground state of a bosonic ladder with an analog of a magnetic field. We vary the strength of the hopping aross the rungs of the ladder, and the interaction between the bosons.  We find three phases shown in Fig. \ref{phasediagram}: (1) A ``saturated chiral current phase" (SCCP), where the density is uniform and
opposite currents flow on each leg. The magnitude of the chiral current is set by strength of the magnetic field
and is independent of the interactions or the inter-leg hopping strength. In this regime the only spontaneous broken symmetry is the $U(1)$ condensate phase.  (2) A ``biased ladder phase" (BLP), where the density is higher on one leg than the other.  The fluid velocity is higher on the lower density leg, so the net current is zero.  In addition to the $U(1)$ condensate phase, this has a spontaneous broken $Z_2$ reflection symmetry.  (3) A ``modulated density phase" (MDP), where the atomic density is modulated along the ladder.  In addition to the $U(1)$ condensate phase, this has a second spontaneous broken $U(1)$ symmetry corresponding to translations of the density wave. We further study the fluctuations of the condensate in the BLP, finding a roton-maxon like excitation spectrum. Decreasing the hopping along the rungs softens the spectrum. As the energy of the ``roton" reaches to zero, the BLP becomes unstable. We describe the experimental signatures of these phases, including the response to changing the frequency of the Raman transition.

The SCCP and MDP were first introduced by Orignac and Giamarchi \cite{Giamarchi2001}, and the experimentalists interpreted their results in terms
of these phases \cite{Bloch2014}. The BLP has not previously been discussed, but as we explain, the experimental data shows hints of it.

\emph{Model ---}
We consider the Hamiltonian of an interacting Bose gas trapped on a two-leg ladder in a uniform magnetic field,
\begin{eqnarray}
\label{ham}
\notag H_0&=&-J\sum_\ell\left(a^\dagger_{\ell+1}a_{\ell}+b^\dagger_{\ell+1}b_{\ell}+H.c.\right)\\
&-&K\sum_\ell\left(a^\dagger_{\ell}b_{\ell}e^{i\ell\phi}+H.c.\right),\\
H_1&=&\frac{g}{2}\sum_\ell\left(a^\dagger_{\ell}a^\dagger_{\ell}a_{\ell}a_{\ell}+
b^\dagger_{\ell}b^\dagger_{\ell}b_{\ell}b_{\ell}\right),
\end{eqnarray}
where $\ell$ corresponds to the positions along the ladder and the bosonic operator $a_\ell$ ($b_\ell$) annihilates a boson on site $\ell$ of the left (right) leg.
The tunneling strength along the legs is $J$, the tunneling strength across the rungs is $K$, 
and the magnetic flux per unit cell is $\phi$. The model was proposed by Atala \emph{et al.} to describe their experiment on trapped Rubidium atoms \cite{Bloch2014}.
The intra-leg hopping $J$ is set by the intensity of the lasers which create their lattice potential.
The inter-leg hopping $K$ is set by the intensity of a second set of lasers which drive a Raman transition that allows hopping between the legs.
The interaction strength $g$ is controlled by modifying the transverse confinement \cite{Cheng2013}. 
In the experiment, there is only a weak trap in the $z$-direction, and $g$ is very small \cite{Bloch2014}. One could also use a Feshbach resonance to tune $g$ \cite{Cheng2010}.

The single-body Hamiltonian $H_0$ is characterized by a 2 by 2 matrix in the momentum space,
\begin{eqnarray}
\label{ham0}
H_0&=&\sum_k {\bm c}^\dagger_k{\mathcal H}(k){\bm c}_k,\\
\label{ham0k}
{\mathcal H}(k)&=&-2J{\rm cos}k\,{\rm cos}\frac{\phi}{2}+2J{\rm sin}k\,{\rm sin}\frac{\phi}{2}\sigma_z-K\sigma_x,
\end{eqnarray}
where ${\bm c}^\dagger_k=\left(a^\dagger_k,b^\dagger_k\right)$ with
$a_k=\frac{1}{\sqrt{L}}\sum_\ell e^{-i(k+\frac{\phi}{2})\ell}a_\ell, b_k=\frac{1}{\sqrt{L}}\sum_\ell e^{-i(k-\frac{\phi}{2})\ell}b_\ell$, and
$\sigma_x,\sigma_z$ are the Pauli matrices, and $L$ is the length of the ladder. Note $k$, $\phi$ and $L$ are dimensionless. This Hamiltonian is readily diagonalized by
\begin{eqnarray}
\label{trans}
\left(\begin{array}{c}
    a_k\\
    b_k\\
    \end{array}\right)
    &=&
    \left(\begin{array}{cc}
    {\rm cos}\frac{\theta_k}{2} & -{\rm sin}\frac{\theta_k}{2}\\
    {\rm sin}\frac{\theta_k}{2} & {\rm cos}\frac{\theta_k}{2}\\
    \end{array}\right)
    \left(\begin{array}{c}
    \alpha_{k}\\
    \beta_{k}\\
    \end{array}\right)
\end{eqnarray}
with ${\rm tan}\theta_k=\frac{-K/J}{2{\rm sin}k\,{\rm sin}\frac{\phi}{2}}$, yielding
$H_0=\sum_k\big(E_+(k)\alpha_k^\dagger \alpha_k+E_-(k)\beta_k^\dagger \beta_k\big)$,
where the two bands are described by $E_\pm(k)=-2J\,{\rm cos}k\,{\rm cos}\frac{\phi}{2}\pm\sqrt{4J^2{\rm sin}^2k\,{\rm sin}^2\frac{\phi}{2}+K^2}$.
For $K\geq2J\,{\rm tan}\frac{\phi}{2}\,{\rm sin}\frac{\phi}{2}$, the lower band $E_-(k)$ has a single minimum at $k=0$. For
$K<2J\,{\rm tan}\frac{\phi}{2}\,{\rm sin}\frac{\phi}{2}$, it has two minima at $k=\pm k_0$, where $\frac{\partial E_-}{\partial k}\big|_{k=\pm k_0}=0$.
We consider the $N$-body variational wavefunction
\begin{eqnarray}
\label{gs}
|G_{k_0}\rangle=\frac{1}{\sqrt{N!}}\left({\rm cos}\gamma\beta_{k_0}^\dagger+{\rm sin}\gamma\beta_{-k_0}^\dagger\right)^N|{\rm vac}\rangle,
\end{eqnarray}
where $|{\rm vac}\rangle$ is the vacuum state and 
$0<\gamma<\pi/2$ for $k_0>0$ and $\gamma=0$ for $k_0=0$.
In the absence of interactions, this is the ground state for any choice of $\gamma$. 
Even infinitesimal interactions, however, can split this degeneracy.

\emph{Current and density ---}
In this section we explore the properties of Eq. (\ref{gs}). In particular we calculate densities and currents, which are experimental observables \cite{Bloch2014}.

To satisfy the continuity equation, we define the net current and the chiral current,
\begin{eqnarray}
\label{current}
\notag J_n&\equiv&\langle G_{k_0}|\sum_k{\bm c}^\dagger_k\frac{\partial{\mathcal H}(k)}{\partial k}{\bm c}_k|G_{k_0}\rangle/N\\
&=&{\rm cos}^2\gamma\left(J_{k_0}^a+J_{k_0}^b\right)+{\rm sin}^2\gamma\left(J_{-k_0}^a+J_{-k_0}^b\right),\\
J_c&\equiv&\langle G_{k_0}|\sum_k{\bm c}^\dagger_k\sigma_z\frac{\partial{\mathcal H}(k)}{\partial k}{\bm c}_k|G_{k_0}\rangle/N\\
&=&{\rm cos}^2\gamma\left(J_{k_0}^a-J_{k_0}^b\right)+{\rm sin}^2\gamma\left(J_{-k_0}^a-J_{-k_0}^b\right),
\end{eqnarray}
where the currents on each leg are
\begin{eqnarray}
J_{k_0}^a&=&2J\,{\rm sin}\left(k_0+\frac{\phi}{2}\right){\rm sin}^2\frac{\theta_{k_0}}{2},\\
J_{k_0}^b&=&2J\,{\rm sin}\left(k_0-\frac{\phi}{2}\right){\rm cos}^2\frac{\theta_{k_0}}{2}.
\end{eqnarray}
Using the equation $\frac{\partial E_-(k)}{\partial k}\big|_{k=\pm k_0}=0$ and the relation ${\rm sin}^2\frac{\theta_{k_0}}{2}={\rm cos}^2\frac{\theta_{-k_0}}{2}$,
one can read off $J_{k_0}^a=J_{-k_0}^a=-J_{k_0}^b=-J_{-k_0}^b$.
This implies the net current always vanishes at equilibrium and the chiral current is independent of $\gamma$:
\begin{eqnarray}
J_n&=&0,\\
\label{chiraleq}
J_c&=&4J\,{\rm sin}\left(k_0+\frac{\phi}{2}\right){\rm sin}^2\frac{\theta_{k_0}}{2}.
\end{eqnarray}
We also define the local density on each leg, 
\begin{eqnarray}
\label{density}
n^a(\ell)&\equiv&\langle G_{k_0}|a_\ell^\dagger a_\ell |G_{k_0}\rangle=\bar n^a+\delta n^a_\ell\\
n^b(\ell)&\equiv&\langle G_{k_0}|b_\ell^\dagger b_\ell |G_{k_0}\rangle=\bar n^b+\delta n^b_\ell,
\end{eqnarray}
where the average density on each is $\bar n^a/\bar n={\rm cos}^2\gamma\,
{\rm sin}^2\frac{\theta_{k_0}}{2}+{\rm sin}^2\gamma\,{\rm cos}^2\frac{\theta_{k_0}}{2}$
and $\bar n^b/\bar n={\rm cos}^2\gamma\,{\rm cos}^2\frac{\theta_{k_0}}{2}+{\rm sin}^2\gamma\,{\rm sin}^2\frac{\theta_{k_0}}{2}$,
where $\bar n=N/L$ is the average density. The density modulations are the same on each leg:
$\delta n^a_\ell/\bar n_a=\delta n^b_\ell/\bar n_b=\frac{1}{2}\,{\rm sin}2\gamma\,{\rm sin}\theta_{k_0}{\rm cos}2k_0\ell$.
Note the modulation is largest at $\gamma=\pi/4$ and vanishes at $\gamma=0$.

\begin{figure}[!htb]
\includegraphics[width=7.5cm]{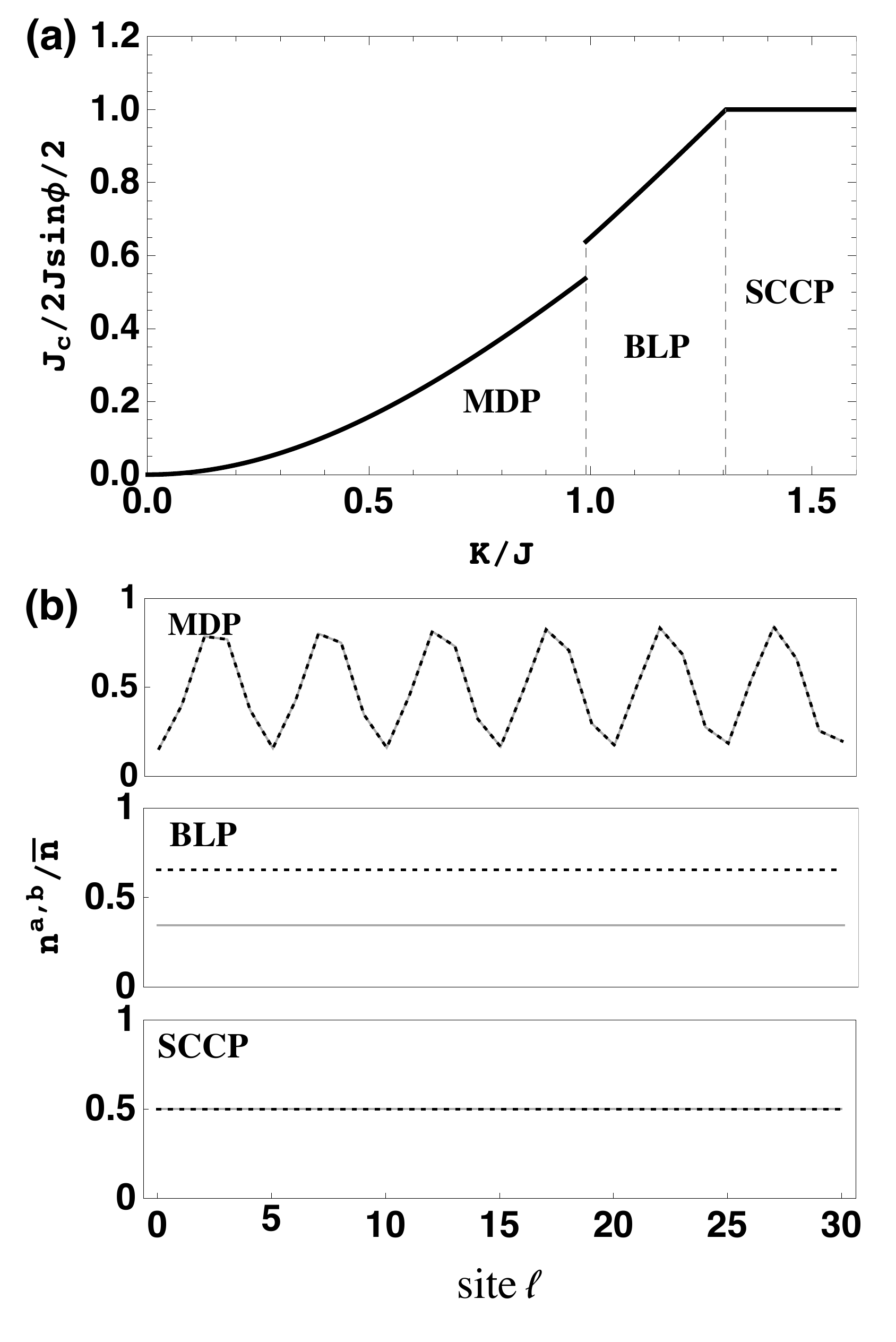}
\caption{Chiral current and atomic density.
(a) Chiral current as a function of tunneling strength $K/J$. The current is discontinuous
at the boundary between the MDP and BLP, indicating a first-order transition, whereas the current is continuous across the
BLP to SCCP boundary. The slope is discontinuous indicating a second-order transition. 
(b) Atomic density as a function of lattice site $\ell$. In the MDP, the density of each leg is equal but modulated along the ladder.
In the BLP, the density is higher on one leg than the other. In the SCCP, the density of each leg is equal and uniform.
For these plots the interaction strength is $g\bar n/J=0.2$ and the magnetic flux is $\phi=\pi/2$.
\label{chiral}}
\end{figure}

\emph{Phase diagram ---} We now consider the interaction term $H_1$. 
Treating Eq. (\ref{gs}) variationally and allowing $k_0$ to be a free parameter, we study the energy
\begin{eqnarray}
\label{energy1}
E(\gamma,k)\equiv\langle G_{k}|H_0+H_1|G_{k}\rangle/N=E_-(k)+E_{\rm int}(k),
\end{eqnarray}
where
\begin{eqnarray}
\label{int}
E_{\rm int}(k)=\frac{g\bar n}{2}\left(\left(\frac{3}{4}{\rm sin}^2\theta_k-\frac{1}{2}\right)
{\rm sin}^22\gamma-\frac{1}{2}{\rm sin}^2\theta_k+1\right).\nonumber\\
\end{eqnarray}
This ansatz describes the three phases in Fig. \ref{phasediagram}.
We minimize $E(\gamma,k)$ with respect to $\gamma$ and $k$.
The only $\gamma$-dependence is in Eq. (\ref{int}).
For $\frac{3}{4}{\rm sin}^2\theta_k-\frac{1}{2}\geq0$, 
the energy minimum is at $\gamma=0$.
For $\frac{3}{4}{\rm sin}^2\theta_k-\frac{1}{2}<0$,
the energy minimum is at $\gamma=\pi/4$.
As can be inferred from the expressions following Eq. (\ref{trans}),
${\rm sin}^2\theta_k=\frac{K^2}{K^2+4J^2{\rm sin}^2k\,{\rm sin}^2\phi/2}$.

For $\gamma=0$, the density is uniform along the ladder, and the chiral current is 
given by Eq. (\ref{chiraleq}), with $\frac{\partial E(\gamma=0,k)}{\partial k}\big|_{k=\pm k_0}=0$.
When $k_0=0$, the density of the each leg is equal, with $n_a=n_b=n_0/2$, and the chiral current is saturated, with $J_c=2J{\rm sin}\frac{\phi}{2}$. 
We call this phase the ``saturated chiral current phase" (SCCP), as shown in Fig. \ref{chiral}.
In the SCCP, the only broken symmetry is the $U(1)$ condensate phase.
For $k_0>0$, the density is higher on one leg than the other, which breaks the $Z_2$ reflection symmetry.
We call this phase the ``biased ladder phase" (BLP). The transition between the BLP and SCCP is second-order,
and as illustrated in Fig. \ref{chiral}(a), the chiral current is continuous across transition. 
Note the BLP has a two-fold degeneracy since the choice of the leg with a higher (lower) density is arbitrary.
In our ansatz, this two-fold degeneracy is associated with symmetry $k_0\rightarrow-k_0$.

For $\gamma=\pi/4$, the density is modulated along the ladder, which supplements the broken $U(1)$ condensate phase,
with a second broken $U(1)$ symmetry: the energy is unchanged if one adds an arbitrary phase to $\beta_{k_0}^\dagger$ or $\beta_{-k_0}^\dagger$ in Eq. (\ref{gs}).
This second $U(1)$ phase is related to translations of the density modulation.
We call this regime the ``modulated density phase" (MDP). The transition between MDP and the former two phases is first-order, as
$\gamma$ changes discontinuously. Furthermore, we see the chiral current has a discontinuous
jump between the MDP and BLP in Fig. \ref{chiral}(a). 
The size of the current jump is determined by the interaction strength $g$, and disappears when $g$ is zero.

Note for $\gamma=\pi/4$, Eq. (\ref{gs}) is a special case of a more generic ansatz
$|T_{k_0}\rangle=\frac{1}{\sqrt{N!}}\left(\sum_n c_n\beta_{nk_0}^\dagger\right)^N|{\rm vac}\rangle$
where $\sum_n|c_n|^2=1$ \cite{Wei2011}. 
Although we do not plot the results, we have studied this more general ansatz. We find very few changes: 
the boundary between the phases is only shifted to a slightly larger tunneling strength $K/J$.
The symmetry of each phase is unchanged. The shift vanishes as $g\rightarrow0$.

\emph{Stability and Roton ---}
We now study the stability of Eq. (\ref{gs}) when $\gamma=0$.
We find the excitation spectrum of the BLP has a maxon-roton like structure.

To calculate the excitation spectrum, we truncate the Hamiltonian to the lowest band 
\begin{eqnarray}
H=\sum_kE_-(k)\beta_k^\dagger\beta_k+\frac{1}{2L}\sum_{kpq}\Gamma_{kpq}\beta_{k+q}^\dagger\beta_{p-q}^\dagger\beta_p\beta_k
\end{eqnarray}
where
\begin{eqnarray}
\notag\Gamma_{kpq}&=&g\bigg({\rm sin}\frac{\theta_{k+q}}{2}{\rm sin}\frac{\theta_{p-q}}{2}{\rm sin}\frac{\theta_{p}}{2}{\rm sin}\frac{\theta_{k}}{2}\\
&+&{\rm cos}\frac{\theta_{k+q}}{2}{\rm cos}\frac{\theta_{p-q}}{2}{\rm cos}\frac{\theta_{p}}{2}{\rm cos}\frac{\theta_{k}}{2}\bigg).
\end{eqnarray}
The ansatz in Eq. (\ref{gs}) with $\gamma=0$ is equivalent to setting $\beta_k=\sqrt{N}\delta_{kk_0}$. We add fluctuations, 
writing $\beta_k=\sqrt{N}\delta_{kk_0}+(1-\delta_{kk_0})\chi_{k-k_0}$. 
To quadratic order in the operators $\chi_k$,
\begin{eqnarray}
\label{quadratic}
\notag \bar H/N&=&E(\gamma=0,k_0)+\sum_{k>0}\zeta(-k)\\
&+&\sum_{k>0}
(\chi_k^\dagger,\chi_{-k})
\left(\begin{array}{cc}
    \zeta(k) & \eta(k)\\
    \eta(k) &  \zeta(-k) \\
    \end{array}\right)
   \left(\begin{array}{cc}
    \chi_k \\
    \chi_{-k}^\dagger \\
    \end{array}\right)
\end{eqnarray}
where
\begin{eqnarray}
\notag\zeta(k)&=&E_-(k+k_0)+2g\bar n\bigg({\rm sin}^2\frac{\theta_{k_0}}{2}{\rm sin}^2\frac{\theta_{k_0+k}}{2}\\
&+&{\rm cos}^2\frac{\theta_{k_0}}{2}{\rm cos}^2\frac{\theta_{k_0+k}}{2}\bigg)-\mu,\\
\notag\eta(k)&=&g\bar n\bigg({\rm sin}^2\frac{\theta_{k_0}}{2}{\rm sin}\frac{\theta_{k_0+k}}{2}{\rm sin}\frac{\theta_{k_0-k}}{2}\\
&+&{\rm cos}^2\frac{\theta_{k_0}}{2}{\rm cos}\frac{\theta_{k_0+k}}{2}{\rm cos}\frac{\theta_{k_0-k}}{2}\bigg),
\end{eqnarray}
where we have subtracted the chemical potential $\mu=E_-(k_0)+g\bar n\left({\rm sin}^4\frac{\theta_{k_0}}{2}+{\rm cos}^4\frac{\theta_{k_0}}{2}\right)$
and defined $\bar H=H-\mu N$.
We perform the Bogoliubov transformation $\chi_k=u \rho_{-k}-v\rho_{k}^\dagger$
and $\chi_{-k}^\dagger= -v\rho_{-k}+u\rho_{k}^\dagger$, where $\rho_k$
is the bosonic quasiparticle and $u^2-v^2=1$.
The Hamiltonian is then diagonalized as
\begin{eqnarray}
\bar H/N=\sum_{k>0}\epsilon_{k}\rho_k^\dagger\rho_k+
\epsilon_{-k}\rho_{-k}^\dagger\rho_{-k}+{\rm const.}
\end{eqnarray}
where the Bogoliubov excitation spectrum is
\begin{eqnarray}
\epsilon_k=\sqrt{\frac{(\zeta(k)+\zeta(-k))^2}{4}-\eta^2(k)}+\frac{\zeta(-k)-\zeta(k)}{2}.
\end{eqnarray}

\begin{figure}[!htb]
\includegraphics[width=8.5cm]{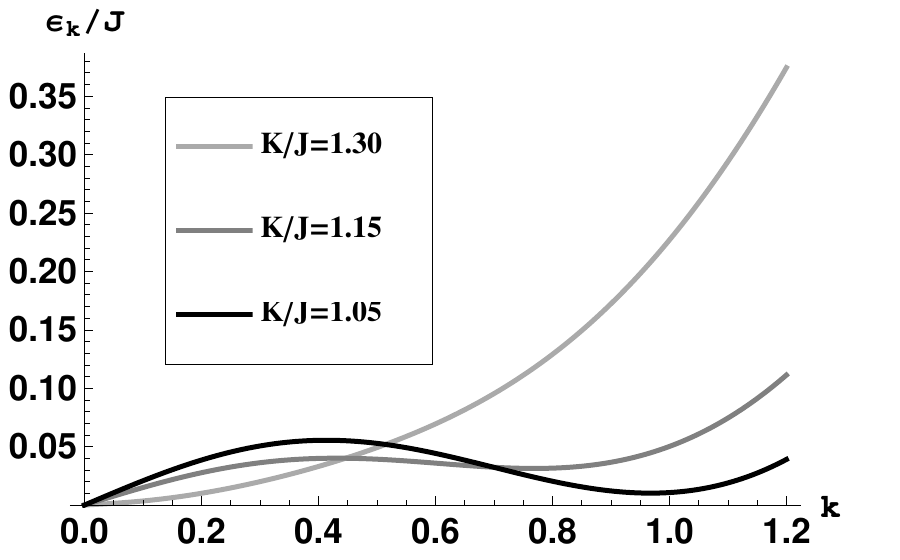}
\caption{
Bogoliubov excitation spectrum $\epsilon_k/J$ for $g\bar n/J=0.2$. The ``maxon-roton" like structure develops as one decreases the 
tunneling strength $K/J$. When the energy of the ``roton" hits zero, the BLP is unstable. This corresponds to a spinodal, and the first-order
thermodynamic BLP-MDP phase transition generically preempts it.
\label{roton}}
\end{figure}

In the BLP, this spectrum has a maxon-roton like structure, as shown in Fig. \ref{roton}. Decreasing the 
tunneling strength $K/J$ softens the spectrum. As the energy of the roton reaches to zero, 
the BLP becomes unstable. This corresponds to a spinodal, and the first-order transition 
between the BLP and MDP generically preempts it.

\emph{Experimental signatures ---}
In this section we describe experimental signatures of these phases.
A local density measurement can distinguish the three phases, as can a measure of local currents.
Some of the phases can be distinguished via time-of-flight measurements. Finally, we argue that a
susceptibility measurement can readily identify the BLP.

\begin{figure}[!htb]
\includegraphics[width=8.5cm]{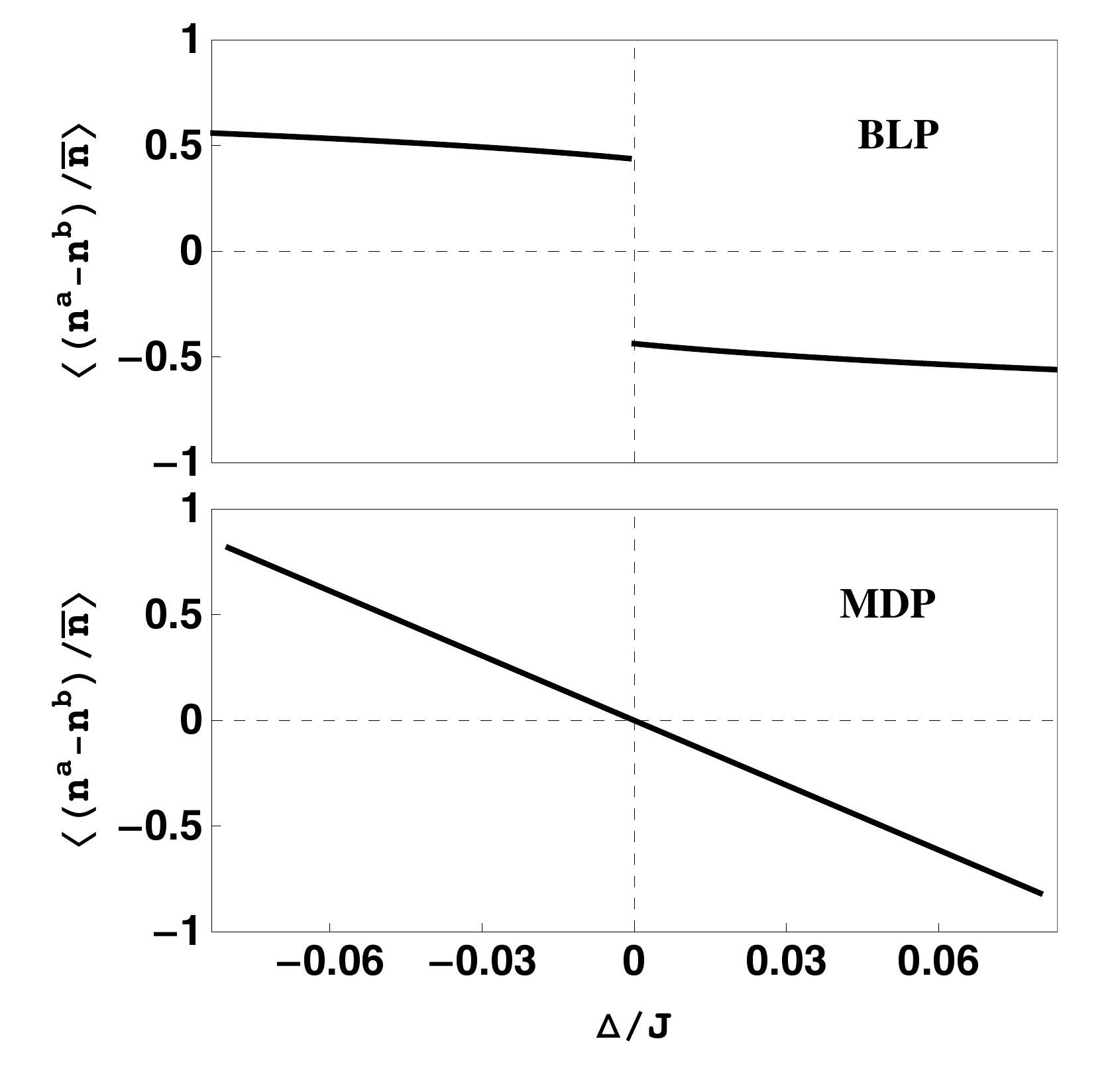}
\caption{
Averaged density asymmetry $\langle(n^a-n^b)/\bar n\rangle$ as a function of the detuning $\Delta/J$. 
The density is calculated by averaging over $30$ sites along the ladder, where we set $g\bar n/J=0.2$,
and $K/J=0.2$ for the MDP and $K/J=1.1$ for the BLP.
\label{detuning}}
\end{figure}

While local density and current measurements can be difficult, the experimentalists in Ref. \cite{Bloch2014}
devised an ingenious surrogate. They isolate each leg of their ladder and further break each leg into a set of dimers.
By looking at the time evolution of this ensemble of isolated dimers, they extract averages of various local correlation functions.
In particular they find that the chiral current saturates for $K/J>\sqrt{2}$. Given their weak interactions, this is consistent with the SCCP
in Fig. \ref{phasediagram}. They also find signatures of spatial inhomogeneities along each leg for $K/J<1$ (see Fig. 4(b) of Ref. \cite{Bloch2014}).
This is consistent with a transition to the MDP. For $1<K/J<\sqrt{2}$, they appear to have a state which is translationally invariant along the ladder,
and has a non-saturated chiral current. This is consistent with the BLP. 
The experimentalists interpreted their data in terms of the SCCP and MDP, which they referred to as the ``Meisner phase" and ``vortex phase". 
They were unaware of the possibility of the BLP, as it has not been previously discussed. 
The experimentalists make a plot of $J_c$ vs $K/J$, similar to Fig. \ref{chiral}(a). While the phase transitions should all be visible
in this graph, the discontinuity between the BLP and MDP vanishes as the interaction parameter $g\rightarrow0$.

Another direct probe of these states is the left-right asymmetry $\delta=n^a-n^b$. In the BLP, $\delta\neq0$.
Unfortunately, the experiment is performed on an array of ladders, and one would expect each ladder to randomly have 
$\delta>0$ or $\delta<0$. The ensemble average will be zero in all phases. 
To avoid this issue, we propose a susceptibility measurement. We envision detuning the Raman lasers from
resonance, which adds to Eq. (\ref{ham}) a term $H_\Delta=\sum_\ell\Delta(a_\ell^\dagger a_\ell-b_\ell^\dagger b_\ell)$.
Such a term can also be engineered by adjusting the geometry of their lattice beams.
In the BLP, any bias $\Delta$, no matter how small, will yield a finite left-right asymmetry. In the MDP or SCCP,
the asymmetry will instead be linear in $\Delta$.  

Figure. \ref{detuning} shows the averaged density asymmetry $\langle(n^a-n^b)/\bar n\rangle$ as a function of the detuning $\Delta/J$
over $30$ sites along one ladder. The discontinuity seen for the BLP can be interpreted as a divergent susceptibility.
In an experiment one would likely see hysteresis in the chiral current for the BLP.
By contrast the MDP has a finite susceptibility.

Finally we consider time-of-flight expansion. In principle one can use this technique to directly measure the momenta of all the particles.
In the SCCP, the atoms on the left legs all have momentum $k_0=\phi/2$ along the ladder, and the atoms on the right legs all have momentum $-k_0$. In the BLP
the characteristic momentum is reduced to $k_0<\phi/2$, but there is still only one momentum peak for each leg.
In the MDP the distribution is bimodal: on each leg there are two different momenta.

To fully interpret time-of-flight images from arrays of ladders, one must take into account inter-ladder coherences. Thus we consider a
more general two-dimensional model with
\begin{eqnarray}
\notag H_0&=&-J\sum_{\ell j}\left(a^{(j)\dagger}_{\ell+1}a_{\ell}^{(j)}+b^{(j)\dagger}_{\ell+1}b_{\ell}^{(j)}+H.c.\right)\\
\notag&-&K\sum_{\ell j}\left(a^{(j)\dagger}_{\ell}b_{\ell}^{(j)}e^{i(\ell+j)\phi}+H.c.\right)\\
&-&\Lambda\sum_{\ell j}\left(e^{-i\lambda}a^{(j+1)\dagger}_{\ell}b_{\ell}^{(j)}e^{i(\ell+j+1)\phi}+H.c.\right),
\end{eqnarray}
where the superscript labels the ladder, and the tunneling strength between adjacent ladders is $\Lambda$.
The phase factors $e^{i(\ell+j)\phi}$ and $e^{i(\ell+j+1)\phi}$ are related to the experimental geometry of the Raman beams, 
and $e^{-i\lambda}$ involves details of the excited state in the Raman transition.
Diagonalizing this Hamiltonian in momentum space, one finds the lower energy band 
$E_-(k_x,k_y)=-2J\,{\rm cos}k_y\,{\rm cos}\frac{\phi}{2}-\sqrt{4J^2\,{\rm sin}^2k_y\,{\rm sin}^2\frac{\phi}{2}+K^2+\Lambda^2+2K\Lambda\,{\cos(k_x+\lambda-\phi)}}$, 
where $k_y$ is the canonical momentum in the $y$-direction (along the leg of the ladder),
and $k_x$ is the canonical momentum in the $x$-direction (perpendicular to the leg of the ladder).
Time-of-flight measures the real momentum, ${\bm p}$, where $a_{\bm p}=a_{{\bm k}-{\bm q}}$ and $b_{\bm p}=b_{{\bm k}+{\bm q}}$
with ${\bm q}=\frac{\phi}{2}(\hat x+\hat y)$.
For completely decoupled ladders, $\Lambda=0$, the energy is independent of $k_x$.
For any finite coupling, $\Lambda>0$, the energy minimum is given by $k_x=\phi-\lambda$. 
We then see that the atoms on the left legs have $p_x=k_x-\frac{\phi}{2}=\frac{\phi}{2}-\lambda$,
and the atoms on the right legs have $p_x=k_x+\frac{\phi}{2}=\frac{3}{2}\phi-\lambda$. Thus atoms from the two legs become spatially separated during time-of-flight.
This spatial structure is seen in Ref. \cite{Bloch2014}.

\emph{Conclusions ---}
We have studied the ground state of a bosonic two-leg ladder in a magnetic field.
We found three phases, corresponding to different types of broken symmetries.
We further studied the fluctuation of the condensate and found a roton-maxon like excitation spectrum. 
Finally, we described the experimental evidence of these phases, and proposed a susceptibility
measurement to further characterize them.

This work is supported by the National Science Foundation Grant no. PHY-1068165.

\end{document}